\documentstyle[12pt]{article}
\addtocounter{footnote}{1}

\begin{document}
\newcommand{\EQ}{\begin{equation}}
\newcommand{\EN}{\end{equation}}
\newcommand{\EQN}{\begin{eqnarray}}
\newcommand{\ENN}{\end{eqnarray}}
\newcommand{\DE}{\Delta}
\newcommand{\EXP}{{\rm exp}}
\newcommand{\E}{{\rm e}}
\newcommand{\I}{{\rm i}}
\newcommand{\A}{\alpha}
\newcommand{\B}{\beta}
\newcommand{\G}{\gamma}
\newcommand{\PA}{\partial}
\newcommand{\ad}{{\rm ad}}
\newcommand{\rank}{{\rm rank}}
\newcommand{\LM}{\Lambda}
\newcommand{\VF}{\varphi}
\newcommand{\th}{\theta_{12}}
\newcommand{\pp}{\partial_{+}}

\newcommand{\mA}{{\cal A}}
\hyphenation{Cou-lomb}
\def\rddots{\mathinner{\mkern1mu\raise1pt\vbox{\kern7pt\hbox{.}}
\mkern2mu
	\raise4pt\hbox{.}\mkern2mu\raise7pt\hbox{.}\mkern1mu}}

\newcommand{\name}{\sc}

\begin{flushright}
YITP/K-978 \\
April 1992

\end{flushright}

\begin{center}
\LARGE
{\bf
Dynamical Topological Phase Transition \\
in \\
Massless Thirring Model
}

\vspace{2cm}
\large
{\name
Hiroshi Nohara
}

\vspace{1cm}
{\it
Yukawa Institute for Theoretical Physics
\\
Kyoto University
\\
Kyoto, 606 Japan
}

\vspace{3cm}
ABSTRACT
\end{center}
We study the topological nature of both
 isotropic and anisotropic SU(N)
Thirring model. It is shown that
 in the isotropic model there exists the
special point where the system lives in the topological phase and that
in the anisotropic one which is obtained by introducing two coupling
constants and has U(1) symmetry, we present a simple mechanism of the
dynamical topological phase transition
 which takes place at the infinite
energy scale.
\baselineskip=0.8cm
\normalsize


\newpage
The chiral massless Thirring model is one of the simple models which
 realize the non
-perturbative effect of the fundamental interaction. In \cite{Gross}
, it is shown that the famous 1/N expansion analysis enables us
to see the non-perturbative effect of the four-fermions coupling,
that is, the dynamical mass generation.
 But, this is not a whole story.
This type of analysis is valid still
 in the small coupling region while
in the strong one it may not be applicable and we have to take into
account a new non-perturbative effect in the theory. For example,
 in the strong coupling
region, it may happen that the higher order correction causes
the symmetry breaking or a drastic change of the phase structure
of the system.

In this paper, we make a primitive observation
 on the topological aspect
of the chiral SU(N) Thirring model. As we will see, there exists the
 special
point where the local internal symmetry is generated, the dynamically
 generated
mass becomes infinite and the model possesses the topological nature.
As a by-product, we present a simple scenario in which the dynamical
phase transition happens at the high energy limit and the system moves
 to
the region of the topological phase. This type of the phase transition
 is
different from the K-T type and gives us a new mechanism for the one
 generated
by a pure dynamical process.

We first consider the SU(N) Thirring model given by the following
 Lagrangian
\EQ
L =  \psi^{\dagger i}_{L}\partial_{R} \psi^{i}_{L}
   + \psi^{\dagger i}_{R}\partial_{L} \psi^{i}_{R}
   + g j^{a}_{L}j^{a}_{R}, \ \ \ \
j^{a}_{L,R} = \psi^{\dagger }_{L,R}T^{a}\psi_{L,R} ,
\EN
where $\psi^{i} (\psi^{\dagger i})$ are Dirac fermions and $T^{a}$ are
 the
generators of U(N). For small value of $g <<1$, the classical
 continuous
U(N) chiral symmetry
\EQ
\psi^{i}_{R} \rightarrow G(z)\psi^{i}_{R},\ \ \
\psi^{i}_{L} \rightarrow \overline{G}(\overline{z})\psi^{i}_{L},\ \ \
G(z), \overline{G}(\overline{z}) \in {\rm U(N)},
\EN
is broken by the non-perturbative quantum effect of four fermions
 interactions
 and the dynamical mass is generated as discussed in \cite{Gross}.

Instead of this fact, let us try to find certain value of $g$ for which
the theory has some type of the continuous internal symmetry in the
Lagrangian through the perturbative approach.

We assume the following continuous non-chiral U(N) transformation
which the classical theory does not share,
\EQ
\psi^{i}_{L,R} \rightarrow G(z, \overline{z})\psi^{i}_{L,R},\ \ \ \
\psi^{\dagger i}_{L,R} \rightarrow \psi^{\dagger i}_{L,R}
G^{-1}(z, \overline{z}),\ \ \
G(z, \overline{z}) \in {\rm U(N)} \label{gaugetrans}.
\EN
Taking it into account that the each term in Lagrangian is regularized
by point-splitting procedure
\EQ
\psi^{\dagger i}_{L}(z,\overline{z})\psi^{j}_{L}(w,\overline{w})
 \sim  \delta_{ij}/(\overline{z}-\overline{w})+ \cdots , \ \ \
\psi^{\dagger i}_{R}(z,\overline{z})\psi^{j}_{R}(w,\overline{w})
 \sim  \delta_{ij}/(z-w) + \cdots ,
\EN
the fermion currents are transformed into
\EQ
j^{a}_{L,R} \rightarrow   j^{b}_{L,R}Tr(G^{-1}T^{b}GT^{a})
                          +Tr(\PA_{L,R}G^{-1}\cdot GT^{a}).
\EN
Substituting these equations into the Lagrangian, we see that by
 setting
$g=1$, the Lagrangian is invariant up to the polynomials of $G
(G^{-1})$ and its derivatives which does not change the dynamics at
all.

In this way, we define the massless Thirring model with the local
 internal
symmetry, that is, the gauge symmetry at the special value of the
coupling constant at the lowest order. For the time being, we assume
without any reason that this symmetry is not broken by the higher order
correction. To see the feature of this model let
 us consider for
example, the correlation function of the fermion currents
 $j^{a}_{L,R}$,
\EQ
F(z_{i})=\langle \prod_{i} j^{a_{i}}_{L,R}(z_{i},\overline{z_{i}})
\rangle_{g=1}.
\EN
This is diverge because of the local symmetry
 and the correlation length
is infinite. From this fact we expect that the phase transition takes
place and the system lives in the ordered phase here.

Let us consider the U(1) Thirring model for illustration. This
model is massless for any value of $g$ in the quantum level and
 by using
the bosonization formula, the Lagrangian can be rewritten as
\EQ
L=(1-g)\PA_{L}\phi \PA_{R}\phi .
\label{Boson}
\EN
At $g=1$, the Lagrangian is zero and the correlation function of
 currents
$\langle \PA_{L,R}\phi (z) \PA_{L,R}\phi \rangle $ becomes diverge.
 It is
instructive to mention the 1 dimensional quantum spin chain
 in this case. The
U(1) Thirring model is obtained by performing the continuum
 limit of the spin 1/2 spin chain
whose Hamiltonian is
\EQ
H=-\sum_{n}S_{n}^{+}S_{n+1}^{-}+S_{n}^{-}S_{n+1}^{+}
+qS_{n}^{z}S_{n+1}^{z},
\label{spin}
\EN
where $n$ is the index of site on the 1 dimensional chain and the
$S_{n}^{i}$ are spin operator in spin 1/2 representation of SU(2)
 on site $n$.
\cite{Aff}
At the interval $ -1<q<1$, the system is critical
 and at the point $q=1$
which corresponds to $g=1$, the state is in the ferroelectric-ordered
phase and completely frozen. Indeed the correlation function $\langle
 S^{z}(x)S^{z}(y)\rangle$ goes to infinity as $q$ approaches unity. It
should be remarked that the system $(\ref{spin})$ at $q=1$ can not
 be described
as the limit $g \rightarrow 1$ in (\ref{Boson}) since the form
of the dispersion relation in the ferroelectric phase is
 not relativistic
one.

The problem is how to define the SU(N) Thirring model at
$g=1$. We choose to construct the relativistic theory way and to this
end formulate the system by introducing external gauge fields
$A_{L,R}^{a}$ and rewrite the Lagrangian as
\EQ
L= \psi_{L}^{\dagger}(\PA_{R}+ eA^{a}_{R}T^{a})\psi_{L}
  +\psi_{R}^{\dagger}(\PA_{L}+ eA^{a}_{L}T^{a})\psi_{R}
  +e^{2} A^{a}_{L}A^{a}_{R} \label{2QCD}.
\EN

This theory is the two dimensional QCD in the strong coupling limit and
there are no kinetic terms of the gauge fields. It is
first discussed in Polyakov's lecture \cite{Pol}
 and he showed that if the gauge group is U(N)
the total current anomaly is canceled out and
 the central charge is equal
to zero. In fact, if we do not fix the gauge,
 the constraint in (\ref{2QCD})
for the total current reads
\EQ
j^{a}_{L,R,total}=j^{a}_{L,R}+eA^{a}_{L,R}=0.
\EN
This equation indicates the analogy between the operators
$j^{a}_{L,R}$ and $-eA^{a}_{L,R}$ for example, in the transformation
 property by $G$. Indeed we have already seen this property
 in (\ref{gaugetrans})
and consider this equation to be valid as the operator equation.
Since the central charge is zero we conclude
 that the massless Thirring
model can be defined to be a relativistic topological theory
 at $g=1$ by
adopting the formulation based on the Lagrangian (\ref{2QCD}).
It happens that the fermion current is transmuted into
 the gauge current
and eats up its own dynamical degree of freedom by tuning
 the coupling constant.
Recently some detail analysis on this model is given by using the BRST
formalism in \cite{Komata}. According to \cite{Komata},
 we can consider
the theory to be massless and define
 the energy momentum tensor which is BRST
exact. Besides the system has the N=2 super conformal
 symmetry by twisting it.
As a relevant topic, the U(1) gauge-Higgs model in two
dimension is studied by using the same method in \cite{Ichinose}.

To justify the above discussion completely we compare our results with
the exact solution obtained through the Bethe-Ansatz approach
 \cite{Andrei}. The exact
solution of the SU(N)(N$\ge$2) Thirring model was given
 by Andrei and Lowenstein
from the modified Bethe Ansatz approach used in the case of
 the famous
Kondo model which describes the dynamics of conduction electrons and
impurities. Let us consider the $n$ particle
state and assume that the size of the system is $l$.
 The cut-off $\Lambda$
of momentum is given by $n/l$ and the continuum limit is obtained
by taking the limit $\Lambda \to \infty$.
It can be shown easily that the Hamiltonian which governs the wave
function in the coordinate space is essentially
equivalent to the one for $n$ particles with the interaction
 of the delta
function type \cite{Andrei},
\EQ
H=-\sum_{i}^{n}\PA_{i}^{2}+c(g)\sum_{i<j}\delta (x_{i}-x_{j}),
\EN
where $c$ is the effective coupling constant which depends
 on $g$ as
$c(g) \sim g/(1-g^{2})$. Since the coupling constant is diverge at
$g=1$, the state can be shown to be frozen in a following way.
Let us consider the path integral
\EQ
Z=\int \prod_{i}{\cal D}x^{i} e^{-\int L}
=\int \prod_{i}{\cal D}x^{i}e^{-c(g) \int L/c(g)},\ \ \
L=-\sum_{i}^{N}\PA_{i}^{2}-c(g)\sum_{i<j}\delta (x_{i}-x_{j}),
\EN
and at $g=1$, only the solution of the equation of motion contributes
to $Z$ and is trivial since in $L/c(g)$ the kinetic terms becomes zero.

We mention briefly the renormalization flow. The special point we have
 discussed is unstable. Indeed, in the region $0<g<1$, the
renormalized coupling constant decreases ($g \rightarrow 0$)
 as discussed
in \cite{Gross} and in $g>1$, it increases ($g \rightarrow \infty$).
 In the
former region, the dynamical mass is generated and in the limit $g=1$ is
diverge as $\sim \Lambda $. In the latter one the theory
 is non-unitary since
the coefficient of the kinetic term of U(1) bosonic
 field $\PA_{L,R}\phi =
\psi^{\dagger i}_{L,R}\psi^{i}_{L,R}$ is negative.

So far we analyzed the ordinary SU(N) Thirring model and
showed that some special feature is realized at the point which is
 located on the
boundary of the unitary region. From this fact, it is tempting to treat
the anisotropic Thirring model which is obtained by introducing
distinct coupling constants and see the topological feature since the
non-unitary region would appear in a more non-trivial way and there
 would
be a special line instead of the point. We will see a simple
 mechanism for
the dynamical topological phase transition on the line.

To construct the model, we introduce two coupling constants $g_{h},
 g_{v}$
and define the Lagrangian as
\EQ
L =  \psi^{\dagger i}_{L}\partial_{R} \psi^{i}_{L}
   + \psi^{\dagger i}_{R}\partial_{L} \psi^{i}_{R}
   + \sum_{a \in \delta^{0}}g_{h} j^{a}_{L}j^{a}_{R}
   + \sum_{a \in \delta^{\pm }}g_{v} j^{a}_{L}j^{a}_{R},
\EN
where $\delta^{0}$ and $\delta^{\pm}$ denote
 the Cartan subalgebra of U(N)
and non-Cartan one respectively.

The special line we will discuss from now on is the one defined
 by $g_{h}=1$
on which there is the local U(1) symmetry $\psi^{i} \rightarrow
e^{i \theta ( z,\overline{z})}\psi^{i} $ in the system. To see this,
note that the interaction part of the abelian currents
 can be rewritten
as
\EQ
g_{h} \sum_{i=1}^{N}
      \psi_{L}^{\dagger i}\psi_{L}^{i} \psi_{R}^{\dagger i}
\psi_{R}^{i},
\EN
and is that of the N-copies of the U(1) Thirring model.
 We remark that at the
points ($g_{h},g_{v}$)=(1,1) (1,0), the system is topological.

Let us set $g_{v}$ very small and assume
 that the perturbative analysis
is sufficient to study a qualitative feature of the model.
By bosonizing the fermion fields $\psi_{i}$, we see that only the
interaction terms between non-abelian currents remain
\EQ
g_{v}(\Lambda /\mu)^2
\sum_{k \ne l }
:
e^{i(\phi_{L}^{k}-\phi_{L}^{l})} e^{i(\phi_{R}^{l}-\phi_{R}^{k})}
:.
\EN
where $\Lambda$ is the ultra violet cut-off and $\mu$ is the scale
parameter. The renormalized coupling constant
$g_{v}^{r}$ vanishes as $\sim 1/ \Lambda^{2}$ and the system goes to
the stable point $(g_{h}, g_{v})=(1,0)$ where
 the system is topological
as we explained before. The absence of the kinetic term implies
that there is no scattering process and the S matrix is unity.
It should be noted that the theory is not topological
 in the sense that
we have the parameter $\Lambda$ and the renormalization flow
 of $g_{v}$.

In the case of SU(2), the exact solution was derived
 through the Bethe-Ansatz
technique in \cite{Truong,Duty}. In the diagram of the renormalization flow,
the
 line is denoted by A and called the Luther-Emery line.(See Fig. 1)
 It can be shown that the
theory is massless and free on this line for $0<g_{v}<1$ at the ultra
violet limit $\Lambda \rightarrow \infty$ and hence the
above discussion is still correct in the strong coupling region.
 For other
case, although the exact solution is not constructed explicitly
 yet and
we can not justify the argument in the case of the strong coupling, we
believe that there exists an line on which the system behaves in a
 similar
way.

Thus we present the simple scenario for the asymptotic topological
theory on the Luther-Emery line while the theory is asymptotic free
on the other line $0<g_{h}=g_{v} <1$. On this line the system has
 the local
U(1) internal symmetry and at the end point, becomes the sum of
the decoupled U(1) Thirring models each of which lives
 in the topological phase.

The lesson we learned in both isotropic and anisotropic
 Thirring model is
that on the renormalization flow, the local internal symmetry
 is preserved and
if the symmetry is enhanced enough to carry away the dynamical
 degree of freedom
at the fixed point, the theory is topological.
 For example, in the case
of the isotropic O(N)(N$>$2) Thirring model, we have
 the same story at certain
value of coupling constant, but in the anisotropic case,
 can not realize the
dynamical phase transition in a similar way
 since the U(1) current can not
exist.

In previous argument we assumed that the number of
 fermion is finite ($N < \infty$).
Now we mention the case of the infinite number of fermions and
consider the SU($\infty$) Thirring model. From the exact solution
it is easily seen that the dynamical mass is infinite for any value
of $g$ and only the U(1) bosonic field exists. As before at $g=1$
 system
has the local U($\infty$) symmetry the generators of
 which constitute
the k=1, U($\infty$) algebra. By rescaling fermion fields as
\EQ
i/N \rightarrow t, \ \ \
\psi^{i} \rightarrow \sqrt{N}\psi^{i} (N \rightarrow \infty ),
\EN
they can be formally considered to live in the three dimensional space
($z, \overline{z}, t$). When we have no dynamics in any direction
 at all
it is permitted to call the system the three dimensional
 topological theory.

Finally we make brief comments. It is not yet clear
 why the topological
aspect appears at the special points. The basic observation is that
the higher order correction which is not observed
 in the 1/N expansion analysis
contribute somehow and the system has
the local internal symmetry there.
It may be necessary to consider this symmetry
from a different point of view. We also hope
 that the mechanism of the dynamical
topological phase transition may give us a key
 to build more interesting
models such as realize a general covariant version.

We thank  Professors T. Yoneya, P.P. Kulish, T. Inami, I. Ichinose,
S. Odake for useful discussion and in particular for N. Kawakami and
J. Suzuki for kindly explaining various subjects on the spin systems.

\end{document}